\documentclass{article}
\usepackage[utf8]{inputenc}
\usepackage{appendix}
\usepackage{hyperref}
\usepackage[round]{natbib}
\usepackage{parskip}
\usepackage{authblk}
\usepackage{ragged2e}
\usepackage[UKenglish]{isodate}
\usepackage[final]{changes}

\title{A review of problem- and team-based methods for teaching statistics in Higher Education}

\author[1]{Elinor Jones}
\author[2,3]{Tom Palmer}
\affil[1]{Department of Statistical Science, University College London, London, UK}
\affil[2]{Department of Mathematics and Statistics, Lancaster University, Lancaster, UK}
\affil[3]{MRC Integrative Epidemiology Unit and Population Health Sciences, Bristol Medical School, University of Bristol, Bristol, UK}

\date{\today}

\begin{document}

\maketitle

\begin{abstract}

The teaching of statistics in higher education in the UK is still largely lecture-based. This is despite recommendations such as those given by the American Statistical Association's GAISE report that more emphasis should be placed on active learning strategies where students take more responsibility for their own learning. One possible model is that of \textit{collaborative learning}, where students learn in groups through carefully crafted `problems', which has long been suggested as a strategy for teaching statistics. 

In this article, we review two specific approaches that fall under the collaborative learning model: problem- and team-based learning. We consider the evidence for changing to this model of teaching in statistics, as well as give practical suggestions on how this could be implemented in typical statistics classes in Higher Education.

\end{abstract}

\section{Introduction}
University courses in statistics have traditionally been given in the instructional style, in which a lecturer transcribes a set of notes for students over a course of lectures. In this process students are passive recipients of information. This method of delivery can be scaled up to cope with ever increasing class sizes, a crucial factor in determining which teaching methods could realistically be implemented, but the quality of the resulting education is questionable. 

The American Statistical Association, however, specifically endorses a more active approach to teaching with students taking responsibility for their own learning \citep{gaise-report}.  Going further, the idea that statistics education should resemble statistics practice - in terms of presenting legitimate and relevant statistical research questions as part of the learning process \citep{rumsey}, and relying on cooperation, communication, and team-work \citep{roseth} - is clearly advantageous  but does not a always happen in higher education. 

Collaborative learning, where students learn in groups through carefully crafted `problems', has long been suggested as a strategy for teaching statistics \citep{garfield_1993, gaise-report, roseth}. Despite recommendations, statistics education in higher education in the UK is still largely lecture-based, though the tide is slowly turning. 

In this article, we review two approaches that fall under the collaborative learning model: problem- and team-based learning (PBL and TBL). We focus on PBL and TBL for two reasons. Firstly, they provide a strategy for fundamentally changing the nature of how statistics is taught throughout a course or module, rather than possibly one-off activities to promote effective learning. Secondly these learning models have been used extensively in other disciplines  to good effect, with a considerable body of evidence documenting their advantages. Though the review has in mind introductory undergraduate statistics, we hope that the ideas discussed here may also be useful to those teaching more advanced courses.

The paper is organised as follows. In Section \ref{group} we describe both PBL and TBL, and consider the evidence for using these strategies - both specifically for statistics and also more generally - in Section \ref{evidence}. In Section \ref{practice} we consider the practicalities of using these methods for the teaching of statistics and offer tips for effective implementation.

\section{Group-based enquiry-driven teaching methods}\label{group}
PBL and TBL have made their mark in a number of disciplines, including medicine and allied health professions, business, and engineering. Both approaches fall under the umbrella of `active learning', loosely defined as engaging students in activity, which have been advocated for STEM disciplines including Mathematics \citep{braun2017does}. In this case, the activity consists of students learning through a sequence of carefully crafted problems in small groups or teams. 

The set-up, and therefore the nature of how students learn, is different. In PBL, the problem posed becomes the source of learning:  students become independent seekers of information in order to provide a solution, but under the guidance of a facilitator. For TBL, however, students learn first by using resources made available by the instructor and class time is then dedicated to applying this knowledge to solve the problem in teams.

In this section we review the `classic' implementation of both PBL and TBL. Examples of possible variations on these are given later in Sections \ref{evidence} and \ref{practice}.

\subsection{Problem-based learning}
Problem based learning has been used in medical schools and law schools as early as the 1960s, for example by McMaster Medical School, and with increasing uptake since the 1980s \citep{knight.yorke,Boud1997,Schwarz2001}. The traditional instructional approach to medical education, consisting of an intensive pre-clinical period of basic science lectures followed by a clinical teaching programme, has been criticised for failing to equip doctors with all the skills they needed and for not providing students with the context of how their knowledge should be applied \citep{pbl.medschool}. 

In the UK the General Medical Council set the requirements for how medical students should be trained. They advocate PBL for the following reasons \citep{pbl.medschool,savery.2006}:

\begin{itemize}
\item students must have responsibility for their own learning, since learning is most effective when it is active;
\item problem scenarios facing students should be complex, since real-world medical problems are rarely straightforward;
\item learning should be integrated from a wide range of disciplines and subject areas;
\item learning should integrate collaboration, since clinical practice demands that doctors share information and work constructively with others;
\item students should share with their work groups what they have learned and how that contributes to the solution of the problem;
\item a summary analysis of what has been learned should be undertaken because reflection and evaluation are critical;
\item self and peer assessment should be regularly undertaken.
\end{itemize}

PBL aims to teach students to identify problems and then to design a set of objectives, the accomplishment of which will lead to the development of the solution \citep{schmidt.1983}. Medicine isn't the only area to widely use PBL: law is the other main area which has adopted this strategy at scale. \citet{Schwarz2001} points out that the challenges faced by a law school have similarites and differences with those faced by a medical school. It is not unreasonable to assume that the same is true in statistics training: many of the aspects listed above apply directly to applied or practical statistics education, with the others requiring only minor modifications in language. Though the focus here is on the teaching and learning of applied, or practical, statistics, there is also evidence that PBL can be used in teaching and learning the more theoretical aspects of scientific disciplines, including  in mathematics \citep{dahl}. There is certainly scope, therefore, to do similarly with teaching more theoretical aspects of statistics. 

\subsubsection{Group formation and nature of problems}
In contrast to traditional lecture courses common in Higher Education, students are randomly assigned to small groups (typically between 6 and 10 students, though groups are changed every few weeks) to work on an open-ended problem, often called scenarios.

Within a PBL group there are specific roles. Each member of the group is expected to peform each role at least once per term. The roles are as follows:

\begin{itemize}
\item Chair -- to move the group through the stages of PBL in a timely manner, to ensure coverage of a topic, and to encourage all of the group to participate.
\item Scribe -- listens and records information (often on a whiteboard), writes up the agreed objectives, and contributes to discussions.
\item Other group members -- contribute to discussions, articulate knowledge, identify strengths and weaknesses in the group's knowledge.
\end{itemize}

\subsubsection{Format of PBL sessions}
The groups meet to discuss the problem, along with a tutor who acts as a facilitator by asking questions and prompts to guide discussion toward the learning outcomes. With each new scenario the students rotate through the different roles (chair person, scribe, group member), which gives them the chance to develop new skills.

During the first group meeting, students identify what parts of their knowledge are lacking in tackling the problem. They then set their own goals in terms of what information they require in order to solve the problem in hand. Each member of the group researches the required information. The group then reconvenes to discuss what they learnt from their self-study, and apply their new knowledge to the problem in hand. There will typically be several days between the first session and the second session.

\subsubsection{Assessment}
Each PBL session is evaluated through surveys in which the students reflect on their learning experiences. The facilitator guides students through this self-assessment of outcomes relative to the goals they set at the start, to show them the extent of their learning. Final assessment can take any form, and need not be reliant on the group-work during PBL. 

\subsection{Team-based learning}

\subsubsection{Team formation and nature of problems}
Similarly to PBL, the fundamental idea of TBL is that students work on professionally relevant problems. That is, problems that are similar to what they might encounter in the workplace \citep{michaelsen}. The learning differs from that of PBL, however. Teams are formed of between five and seven students, but are not randomly allocated. Instead, the instructor carefully creates groups to ensure that they are heterogeneous, for example, in terms of preparation or previous experience. Students do not change groups during the course. TBL is now popular in Nursing and Medical schools \citep{liu-2017}, and though in these settings the focus tends to be on developing applied  knowledge, there are examples of its use in teaching more theoretical aspects of mathematics \citep{parappilly_woodman_randhawa_2019} and  physics \citep{parappilly_schmidt_ritter_2015}. 

\subsubsection{Format of TBL sessions}
\paragraph{Before the session}
Prior to the teams meeting to discuss the allocated problem, each student must prepare for the group work by studying the provided materials. This could be in the form of reading, watching videos, or any other activity that prepares students sufficiently for the task ahead.

\paragraph{During the session -- readiness assurance}
The Readiness Assurance Process (RAP) aims to ensure that all students have the pre-requiste knowledge of the course material self-studied, in order to take part in the later problem solving exercise. 

Each student completes an individual test (individual Readiness Assurance Test, or iRAT), designed to highlight any deficiencies in the student's understanding of the pre-class material. These tests are typically quick-fire multiple choice tests, done using electronic voting equipment so that they are marked instantaneously. 

Once completed, students re-take the test, but this time in their allocated groups (team Readiness Assurance Test, or tRAT). Discussion of the questions among team members is encouraged, so that the tRAT itself becomes a learning tool where students learn from each other. 

Results are available immediately after the tRAT, allowing students to assess their understanding, but also to contest the questions and/or answers. Teams provide written justification as to why they think they deserve a higher mark, with evidence, e.g. from course material, for the instructor to consider. If the instructor finds in favour of the team, additional marks are awarded. However, only teams who contest will be eligible for additional marks, even if the problem detected was common for all groups. This encourages students to question the material, and also helps with team cohesion.

\paragraph{During the session -- lecture}
Finally, a lecture component directed at all groups gives the instructor opportunity to clarify misunderstandings and common conceptual errors which were picked up during the iRAT and tRAT. 

\paragraph{During the session -- working on a problem}
The remaining time is spent on solving a problem using the material learnt. The problems are aimed at testing students' deeper understanding of the course material, while the RAP process tests base knowledge \citep{liu-2017}. Such problems generally satisfy four criteria (commonly known as the ``4S''):
\begin{itemize}
\item the problem must be Significant;
\item the teams have a Specific set of possible answers from which they choose one;
\item each team works on the Same problem;
\item teams must Simultaneously report their final answer.
\end{itemize}

Importantly, the problem must not be easily segmented into smaller parts that different team members can tackle: the idea is that the group works together on the whole problem. 

\subsubsection{Assessment}
Assessment for the course is generally a combination of individual tests (iRAT scores and final exam mark), and groupwork mark (tRAT scores, scores from the problem and peer evaluations). Further summative assessment can take any format.

\section{Evidence of effectiveness}\label{evidence}
Among their recommendations, the GAISE College Report suggests that modern statistics education should teach statistical thinking as an investigative process of problem solving and decision making, should integrate real data with a context and purpose, and foster active learning \citep{gaise-report}. All these attributes are fundamental to both problem- and team- based learning. The GAISE College Report also calls for using technology to explore concepts and analyse data. While in medicine, for example, hands-on patient-based activities may not be possible in a classroom setting,  we can incorporate practical data analysis into problem- or team-based learning in statistics.

The general approach of group-based learning aligns with the constructivist philosophy of learning, where students actively construct their own knowledge rather than passively receiving it \citep{garfield_1993}. Not only do students learn the subject matter in this way, but they develop softer skills in problem solving that captures some of the non-formal learning that happens in the workplace \citep{Eraut2000}. 

Group-based learning is posited as a largely positive strategy for teaching non-specialist students. Though we found no published literature on the effectiveness of such strategies in teaching statistics to specialist students (i.e. those pursuing degrees in the mathematical sciences), group-based learning has been successfully implemented for mathematics students in discrete mathematics \citep{paterson}. 

Though there are numerous approaches to measuring effectiveness in teaching, the evidence relating to group-based learning tend to fall into three categories:
\begin{itemize}
\item performance on end-of-module assessments or similar;
\item long-term retention of information;
\item student enjoyment or engagement with the material.
\end{itemize}

We discuss the findings of other studies in implementing variants of TBL or PBL in each of these categories. Evidence of impact on staff is considered separately in Section \ref{practice}.

\subsection{Performance on end-of-module assessments}
\cite{kalaian}'s meta-analysis of the effectiveness of group-based learning in statistics, in comparison to lecture-based instruction, revealed that their effectiveness is dependent on the type of group-based learning implemented. In particular, cooperative or collaborative learning (for example TBL) was found to be effective while no evidence of improved academic achievement was found for inquiry-based methods (such as PBL).

Though the meta-analysis did not point to an \textit{overall} benefit to using PBL in comparison to lecture-based instruction, there are examples of superior student performance on statistics assessments after a PBL-type course rather than a lecture course \citep{karpiak}. However, it is not clear whether this is genuinely due to better understanding of the course material or some other factors \citep{gijbels,karpiak}. For non-statistics major courses in particular, the use of PBL may be helpful because it generates a constant use for the statistical methodology, and hence provides students with a motivation to learn \citep{Jaki2009a}. Better performance on module assessments could also be a consequence of students engaged in active learning as opposed to learning passively, rather than the effect of the PBL itself, or unwittingly increasing the amount of guidance from PBL tutors to students especially since students benefit from guidance in very small groups \citep{bude}. 

Improved grades on end-of-module tests was also observed for TBL for service-type courses mathematics \citep{nanes} and specifically in statistics \citep{liu-2017,haidet}.

\subsection{Long-term retention of knowledge}
While there is a growing body of evidence to suggest various group-based learning methods improve end-of-module assessments, far fewer studies have looked at the long-term impact of these strategies on knowledge retention. 

We found no studies looking at long-term retention of knowledge and skills in statistics, and only one in teaching medical students \citep{emke}. In this study, which looked at short- and long-term retention of knowledge and compared a cohort of students taught via TBL with a cohort that was traditionally taught, there was some evidence that the TBL group performed better on assessments in the short-term but no evidence that they retained more knowledge longer term.

\subsection{Student enjoyment and engagement}
A large body of evidence in the literature points to group-based learning as being a positive experience for students. That this is an aspect that receives most attention is not surprising given the difficulties in comparing understanding of course material between cohorts. 

Students are generally positive about TBL in mathematics \citep{nanes,krogstie} and in statistics \citep{stclair}. In particular, some reported students finding mathematical ideas more accessible when the material was taught as a TBL class as opposed to traditional lectures \citep{patersonsheryn}. Balancing this overwhelming positivity are some interesting student insights from other studies. Naturally, not all students will enjoy an active group-learning environment \citep{haidet}, but more specifically a group environment can encourage some students to `coast' in TBL maths classes, relying on their team-mates for back-up \citep{patersonsheryn}. Others - perhaps weaker students - may find the team environment intimidating \citep{stclair}. In the latter case however, team-working and communication is an essential skill which should be developed alongside mathematical or statistical skills \citep{nanes,tinungki}.

Though most of the research we found on student engagement was based on TBL, aspects of problem-based learning for large cohorts have been considered.  \citet{klegeris} found that PBL sessions for a pharmaceutical course increased attendance in comparison to traditional lectures. This was trialled with and without student additional marks for attendance. They found that offering such a reward for attendance did not significantly affect attendance rates. For large statistics classes in particular, \citet{bude} found that more guidance from tutors/facilitators during the session resulted in better student perception of the course. They warn that increasing the amount of guidance from tutors in a PBL setting could inadvertently lead students to become passive about their learning and less motivated, though they did not find evidence of this in their study.

 While the evidence on balance suggests improved student engagement through the use of TBL and PBL, it is not clear whether these approaches to teaching will suit all students. Making learning inclusive, for example to those with additional educational needs, may mean that adaptations to PBL and TBL are necessary though to our knowledge there are no published papers exploring this particular aspect. 

\section{Problem- and team-based learning of statistics in practice}\label{practice}
Statistics is perhaps an obvious candidate for group-based learning, rich with opportunities in tackling `real' problems and can easily be framed as a believable and relevant problem for either team- or problem-based learning strategies. It is therefore not surprising that PBL for example has been used in statistics courses for over 20 years \citep{hilmer.1996,boyle.1999}. 

The nature of statistics means it is rather dependent on the order that the material is introduced. Its highly structured and sometimes abstract nature makes teaching statistics via group-based learning a challenge: deficiencies in understanding of basic concepts may cause difficulties in understanding more complex procedures \citep{bland,bude}. Students can't front-load a large amount of information so adaptations may be necessary. \citet{nanes} for example, in teaching a course on linear algebra via TBL, suggests increasing the amount of testing and making the real world problems shorter though care must be taken  not to `teach to the test'. It is not unreasonable to assume that the same issue could arise in teaching statistics in this manner. 

Implementing group-based learning in statistics therefore needs careful consideration, and in some cases modifications may be necessary. In this section we review the major components of group-based learning and give advice on practical solutions to potential issues.

\subsection{The real-world problems}
Problems for group-based learning can take any format, though will be different in nature for PBL and TBL. In TBL, students are required to complete a set of learning activities for the session - such as pre-reading, watching videos, or completing other tasks - when technical information can be conveyed which is relevant in solving the problem. In this way, common procedures such as hypothesis testing and statistical modelling can be taught. Contrast this with PBL: it is probably unrealistic to expect students to tackle data-driven problems solely through PBL \citep{bland}. For example, expecting students to come to the conclusion that a \textit{t}-test is appropriate without some prior knowledge is unreasonable, though introducing these concepts in other ways is possible. Here we discuss alternatives to data-driven real-world problems, and their suitability for both TBL and PBL.

Real-world problems that don’t require directly handling data may be easier to implement in class, especially if computing power is not required. These can still provide a rich learning experience, and indeed may enhance a student’s broad understanding of the subject while alleviating some of the difficulties in obtaining real, relevant, and well structured data for teaching purposes. 

Alternatives to data driven problems could  be of the form of a research paper or similar \citep{bland}. Asking students to read, digest and report back on findings from a research paper - especially if it is of direct interest to the students -  would broaden the scope of statistics education, taking the emphasis away from mechanical details to interpretation of results, and also motivating students to see the power of statistics in their own discipline. This strategy in particular is suitable for both PBL and TBL. In PBL, the task could be phrased around understanding the statistical methods employed in a paper and why they were used. In TBL, students could critically appraise the use of techniques in context and suggest alternative ways of addressing the paper's research questions and/or put forward a different analysis plan for the data collected.

In the same vein, students could be asked to provide advice on a consultancy basis either on the design of an experiment or on the analysis of previously-collected data. For the latter, carrying out the data analysis could be set as a task outside the class (in the case of PBL, for example, before the next group meeting), or even as pre-work in TBL before the next session. Results of which could then be used as a springboard for the following workshop either in terms of discussing output or applying the results to a connected problem. 

These are relatively easy ideas to implement in introductory or even intermediate courses in statistics. Teaching more advanced mathematical statistics in the same vein requires more thought. Embedding the topic of interest into a real-world problem may require a little flexibility in what we think of as `real-world' as has been done in more advanced mathematics courses \citep{nanes}, though this is not always the case (for example, it is not difficult to think of many applications of the central limit theorem). With PBL we have the advantage of a dedicated facilitator who can help to guide students through possibly abstract ideas, and in TBL the course material students read before the group session can provide the necessary theory before tackling the problem. In teaching more abstract concepts like this, teachers may need to provide students with more guidance on how to tackle the problem, especially in the context of TBL, for example in explicitly asking for students to think about designing simulations in order to reach a solution.

Whatever the format of a real-world problem, \citet{garfield_1993} emphasises that the hallmarks of good group activities include that all students contribute to the task in hand, and suggest that this could be done by simply emphasising this. Both PBL and TBL benefit from having problems to solve that cannot be split into smaller sub-problems to be tackled individually.

In Appendices \ref{sec:pblexample} and \ref{sec:tblexample} we provide an outline of a PBL and TBL session, respectively. The PBL session frames a general question about mental health and requires the students to identify the gaps in their knowledge and, somewhat independently, fill those gaps in order to complete the task. The TBL session is based on the more technical area of probability. Here, the pre-session material that students are required to work on ensure that they have the necessary basic understanding of probability which can then be applied to a problem concerning the sensitivity and specificity of medical tests.

\subsection{Teaching space}
Any form of group-based learning benefits from suitable classroom-like teaching space where students can comfortably work in groups. Traditionally for PBL, this requires sourcing a suitable room for each group and having access to learning spaces conducive to group work has been found to improve session outcomes \citep{Schwarz2001, Jones1988}. This is often too complex to manage, especially with ever increasing class sizes in statistics, with the only viable alternative to host sessions in large lecture theatres \citep{nicoll, klegeris, roberts}. With some organisation, however, running PBL in these spaces is not insurmountable.

TBL, by its very nature, isn't hampered by such space constraints and is designed to work in lecture theatres. Not all lecture theatres are created equal, however: single level lecture theatres will make it far easier for students to interact within their team in comparison to the usual sloping tiered theatre. It is not surprising that \citet{espey} found that student attitudes toward team-based learning improved with when students perceived the environment they were in to be a comfortable space in which to work in their teams.

\citet{nicoll} suggests using a classroom that is larger than you need for your group size, to create a more comfortable environment for students and to allow the instructor easy access to each group. Of course, computer labs may be required for problems requiring a data-driven solution. Computer rooms are often easier to set up for groups to work together in the sense that they allow some flexibility in rearranging seating easily to suit each team. It may be better for group cohesion if students are \textit{not} allocated a PC each; one PC per team goes some way to ensure that the students in a group interact with one another rather than each student `doing their own thing'.

\subsection{Staff resources}
Traditional PBL is staff-intensive, requiring a tutor or facilitator for each group. This is unlikely to be an option for many courses, especially as classes in statistics are rapidly increasing in size. Though TBL may seem more practical  as it does not require a facilitator for each group, some institutions have been successful in running PBL sessions with only one facilitator for the entire class. Researchers found that running PBL alongside traditional lectures in biochemistry and physiology, without having a dedicated tutor for each group, was successful in terms of improving problem solving skills as well as student satisfaction and motivation \citep{klegeris}. \citet{nicoll} suggest using on-line platforms such as Poll Everywhere or Twitter so that students can send questions to the lone instructor, who in turn can either project answers for the whole class to see or initiate a class discussion. Without a tutor for each group, however, students need to have some background knowledge of the topic under consideration \citep{nicoll}.

In contrast, \citet{roberts} compares traditional PBL for undergraduate medics with a modification where students tackle PBL-like tasks without a dedicated group tutor. They conclude that the modification is a useful alternative when insufficient staff resources are available. They do find, however, that students with a dedicated facilitator are more likely to perceive the learning activity as being superior though no difference was detected between the two groups in terms of achievement.

\subsection{Creation of groups and student engagement}
 Teaching in the mathematical sciences is often  in traditional lectures with individual assignments and assessments. This is at odds with the nature of mathematics at research level: a fundamentally collaborative endeavour. In statistics, courses with group-work components have been common for quite some time, as reported by \cite{garfield_1993}, \cite{hilmer.1996}, \cite{boyle.1999}, and \cite{Jaki2009}. However, students' experience of this way of working needs to be taken into account at the start of any course using group-based learning. 

Like any new intervention, it may take time for students to get used to the idea of working in groups. Students may engage more with the process once they get used to it, so doing this every now and again might not show the real potential of team-based learning.

In the first instance, explaining the structure of each session, making clear how groups are expected to work together, what is expected from students, how to access help, the role of any facilitators, and general code of conduct, should be the first priority; this is especially so for implementations in large classes \citep{roberts}. In particular, students who have little or no experience with small-group learning strategies like PBL or TBL will need more support, and all sources of help need to be highlighted. All groups need to feel that they understand the task in hand, feel confident that they can speak to a tutor when they need guidance, and that they have sufficient resources. For the latter in particular, this may mean suitable written and/or videoed material. It has been suggested that recording any lecture components of courses - which occur in both PBL and TBL - benefits students \citep{Jaki2009, Jaki2009a}.

Course leaders must be prepared for initial student resistance, especially if students' other courses are taught traditionally. Some students may see group-based learning as a glorified version of self-study (in which case, why pay for an education?) while others may worry that their marks will be unfavourably influenced if having to rely on teammates. Responses to such criticisms and concerns could for example include the pedagogical reasons for teaching statistics in a group-based learning environment, or the benefit in terms of development of soft skills valued by employers. Students worried that their grades will be unfavourably influenced may be placated if they are reassured of the procedures in place to ensure that marks are allocated fairly. Even the strongest students benefit from group-based learning: strong students in groups that work well (e.g. where students are invested in the group's achievement), could benefit from thinking about concepts at a deeper level in order to explain them to weaker members of the group. There are also opportunities for students to assess each other. Strategies such as group members having to assess and provide feedback to each other can help students feel that contribution is rewarded while coasting in the group has negative consequences \citep{freeman_mckenzie_2002}.

How groups are formed can influence the success of a group-based learning course: these learning strategies work only when students engage, and if students are inexperienced in group working then this needs to be monitored and managed carefully \citep{hansen_2006}. In their original format, both PBL and TBL groups are chosen by the course leader and these groups remain together for more than one session to encourage cohesion and ensure diversity of groups. Groups that remain together over a period of time instead of changing on a regular basis tend to display a more positive group dynamic \citep{sweet_michaelsen_2007}. What is more, better student engagement within groups has been noted when groups not only stay together but also work together on a regular basis \citep{theobald_2017}.

Once groups are formed, internal dynamics can influence student performance. Factors that have been found to lower student achievement include being in a group where one student dominates, and/or feeling uncomfortable in the group; these tends to be more prominent factors when the group-work involves high-stakes assessment \citep{theobald_2017}.  

Strategies to facilitate positive group working methods and thus increase engagement may be useful, especially if students are not used to group-based working. One approach that has been suggested to increase students' comfort in groups is to establish group `norms' \citep{theobald_2017}. For example, groups could be required to write and submit their own contract for code of conduct, goals, and methods of working,  which could help in establishing trust through clarifying commitments to each other \citep{hunsaker2011}. These contracts could also be helpful in giving groups a way of dealing with dominant students. If group-work leads to a summative assessment, a contract could in addition allow the group to negotiate mark allocation, for example in how marks are distributed between group- and individual- components, or how individual students will be assessed if peer-assessment is to be used.

It may be tempting to allow students to choose their own groups for a number of reasons, including making students feel more comfortable \citep{theobald_2017}, potentially decreasing student resistance, and ease of administration. In addition, allowing for changing teams in each session may also be tempting. This is especially so if students are not required to attend lecture sessions making steady teams difficult to manage, though this has been shown not to be as effective as groups repeatedly working together \citep{sweet_michaelsen_2007}.

 Allowing students to choose may also compromise the heterogeneity within groups. Moreover, those who don't have an immediate friendship group in class may be severely disadvantaged  when students are permitted to choose their own group: though no published work could be found looking at the effects of this, the authors' own experience  is that students joining a group of students who already know each other may result in problems with group attachment, while creating extra groups consisting of these students may lead to feelings of resentment. 

A half-way house is to involve students in the formation of groups in the sense that they decide on \textit{how} the groups are chosen even though, ultimately, the groups are chosen by the course leader. For example, involving students in decisions around the composition of groups: should they be randomly assigned, how long should groups work together for, should groups be mixed in terms of achievement in previous courses, should groups be balanced in terms of academic background of students, should groups be balanced in terms of gender or any other characteristic? Students who feel that they have some say over how their education is managed are more likely to engage in the first place \citep{bovill_2019}.

\subsection{Staff reaction}
There is evidence to suggest that lecturers find the use of group-based learning a satisfying experience \citep{Jones1988}, and it is reasonable to think that this is because the process is a more interactive experience than didactic teaching. Through this interaction the course leader may naturally find that they have a better understanding of a student's strengths and weaknesses, enabling them to address these issues directly. 

When making the transition to group-based learning \citet{Boud1997} and \citet{Schwarz2001} found that introducing students and faculty members into the new curriculum, as opposed to simply starting it without introductory sessions, helped in its successful adoption.

\section{Discussion}
The importance of quality of teaching in the UK Higher Education sector is emphasised by the introduction of the Teaching Excellence and Student Outcomes Framework (TEF, \citet{tef}) to sit alongside the Research Excellence Framework (REF). The first TEF awards were assigned in 2018, and were evaluated for each University as a whole. The second round of TEF awards, planned for 2021, will include subject/departmental specific assessment. Universities are encouraging teaching staff to modernise their teaching, with focus on the TEF but also the National Student Survey results \citep{richardson}.

There is mounting evidence that traditional lecture courses are not as effective as `active'-type learning strategies in Science, Technology, Engineering, and Mathematics (STEM) subjects \citep{Freeman2014}, and indeed specific evidence that PBL and TBL - as active learning methods - are effective. \citet{bland} goes as far as saying that not using such methods (PBL in this specific case) for statistics and research methods training is detrimental to students, while \citet{tinungki}  highlights the importance of communication in learning mathematics which is well addressed in group-based learning. Indeed, both PBL and TBL are ideally placed to meet the needs of employers, who have often identified poor team working skills, poor written communication skills, and poor oral presentation skills in graduates \citep{knight.yorke}.  This is also identified in the guidelines for undergraduate programmes in the closely allied discipline of data science, which recommend that `projects involving group analysis and presentation should be common throughout the curriculum’ \citep{DeVeaux2017}. Though PBL or TBL are not the only methods for implementing this, similar group based methods are becoming popular in data science education, see for example \cite{cetinkaya-rundel_ellison_2020,  SaltzHeckm2016ch}. 

There has been an explosion of technological advances since \citet{gelman.book} outlined approaches for teaching statistics. In the modern teaching environment both teachers and students are surrounded by resources which weren't previously available. From a student's perspective getting information has never been so easy, speeding up  tasks such as the research component in PBL. From a teacher's perspective, numerous platforms (listed in Appendix \ref{sec:technologies}) make learning and interaction with a class more manageable, whilst student monitoring becomes ever easier with tracking via virtual learning environments or automated marking of online quizzes. These factors contribute to the success of group-based learning strategies.

However, while for statistics modules where there is an applied or practical component there is clear scope to apply group-based learning for the whole or at least part of the module, it is not clear how, or even whether, such learning strategies are suitable for statistics modules which are of a more mathematical nature. In the first instance, the technical nature may make it difficult to create a truly `real-world' problem, and this was also noted by \citet{nanes}. Secondly, highly mathematical modules generally rely more heavily on students' prior knowledge. For students whose prior knowledge isn't strong, personality and motivation is likely to play a large part in their success on the module: a group-based learning module could be intimidating, or the group environment may be the key to success. 

\citet{patersonsheryn} note that few mathematics lecturers use a group-based learning approach to their teaching. One possible reason for this is that rewriting existing courses to entirely group-based learning modules in one fell swoop may not be practical. It needn't be all-or-nothing, however. Introducing students to group-based learning slowly may be beneficial \citep{Boud1997, Schwarz2001}. Elements of group-based learning could be weaved through modules, for example particular topics within a module could be taught in a group-based setting, or even just particular sessions. Care needs to be taken, however, in ensuring that students know why you are doing this, and how it benefits them, otherwise the risk is that students won't engage.

Using group-based learning needn't mean using only PBL, or only TBL, however. Some educators have experimented with combining parts of PBL and TBL to maximise the benefits to students. For example, combining the peer feedback (TBL) with an initial group discussion before the pre-reading assignments (PBL), are possibly positive enhancements to any group learning strategies \citep{dolmans}. Online variants of PBL have also been trialled successfully \citep{jong}. That modifications to the traditional PBL and TBL methods have been successful shows that these strategies are ripe for shaping to fit both the practical constraints of the course, as well as the course content. 

We have shown that implementation of PBL, TBL, or a variant thereof, is possible in the teaching of applied statistics. However group-based learning is implemented, the emphasis on a more rounded student education is clear. Of course, these are not the only active learning strategies. Which is the most effective is the subject of its own debate, and students with different learning styles may prefer different teaching methods \citep{bloom}, but this review shows that there is scope for group-based learning in statistics.

\section*{Acknowledgements}
TP would like to thank Simon Allan and Dr Anne-Marie Houghton from Lancaster University's Postgraduate Certificate in Academic Practice programme for helpful advice. TP was supported by the Integrative Epidemiology Unit, which receives funding from the UK Medical Research Council and the University of Bristol (MC\_UU\_00011/1 and MC\_UU\_00011/3).

The authors would like to thank the two anonymous referees for their detailed and constructive comments which improved the paper.

This pre-print has been published online: Jones E and Palmer T. A review of group-based methods for teaching statistics in higher education. Teaching Mathematics and its Applications: An International Journal of the IMA. Published online 09-03-2021, \url{https://doi.org/10.1093/teamat/hrab002}.

\section*{Author biographies}
Elinor Jones is an Associate Professor (Teaching) in the Department of Statistical Science at University College London. She was awarded a PhD in Statistics and Probability from The University of Manchester in 2009. She is interested in how to engage students in the learning of statistics, particularly through active learning strategies.

Tom Palmer is a Senior Lecturer in Biostatistics in the MRC Integrative Epidemiology Unit, in Bristol Medical School. He was awarded a PhD in Medical Statistics and Genetic Epidemiology from the University of Leicester in 2009. He is interested in different teaching methods and how to apply these to teaching statistics.

\section*{Author contact details}
Elinor Jones, Department of Statistical Science, University College London, Gower Street, London, WC1E 6BT. Email: elinor.jones@ucl.ac.uk

Tom Palmer, MRC Integrative Epidemiology Unit, University of Bristol, Oakfield House, Oakfield Grove, Bristol, BS8 2BN. Email: tom.palmer@bristol.ac.uk. Some of this paper was prepared whilst TP was employed in the Department of Mathematics and Statistics, Lancaster University, Lancaster, LA1 4YF.

\raggedright
\bibliography{bibliography}
\justify

\appendix

\section{Example statistics PBL session: Student mental health -- is there a crisis?}\label{sec:pblexample}

\subsection{Scenario}
Consider the scenario in which you are working as a statistician in the civil service in the Department for Education. Recently there have been several cases of students committing suicide. For example, six students committed suicide in the 2016/2017 academic year in Bristol \citep{mann-2017}.  

Ministers want to know if there really is a crisis in terms of the number of students suffering from mental health problems. You are tasked with investigating this issue.

One minister has read some epidemiological research and is curious as to whether this year's cases are reflective of a truly increasing trend in mental health cases or whether this increase is an anomalous spike in the data.

You are tasked with preparing a short structured report (no more than 5 pages) and presentation on this question.

One issue to consider is that there is limited extra government funding for your analysis: you are not be able to carry out a new study to investigate the issue. Therefore, you should address how you can overcome this limitation.

\subsection{Indicative learning objectives}
Statistical methods:

\begin{itemize}
\item research how to estimate different measures of association;
\item investigate time series methods.
\end{itemize}

Applied statistics:

\begin{itemize}
\item research epidemiological concepts such as incidence and prevalence;
\item understand the differences between different absolute and relative measures of association such as the risk difference and the risk ratio. Consider which measures might be more informative for public health policy makers.
\end{itemize}

Statistical programming:

\begin{itemize}
\item demonstrate how to access publicly available datasets and prepare these for analysis using a software of your choice;
\item show how to present complex longitudinal analyses graphically.
\end{itemize}

\subsection{For lecturers}
This session is designed as a PBL exercise to be run over a week. It is aimed at postgraduate students.

A plan for the sessions could be as follows:
\begin{itemize}
    \item Monday: read and digest problem sheet with brain storming session to identify what issues they will need to research.
    \item Tuesday: Students perform research.
    \item Wednesday: Catch up session - the students assess their progress in their PBL groups, and discuss what extra topics  they need to research.
    \item Thursday: Students finish research and prepare their presentation for Friday.
    \item Friday: Students present their findings (as a group) to the whole cohort of PBL groups.
\end{itemize}

\section{Example statistics TBL session: conditional probability and diagnostic tests}\label{sec:tblexample}

\subsection{Learning objectives}
By the end of this section, students should be able to: 
\begin{itemize}
\item Explain the difference between the union of two (or more) events and their intersection.
\item Calculate the probability of the union of two (or more) events and the intersection of two (or more) events.
\item Distinguish between independent and dependent events.
\item Explain intuitively the idea behind conditional probability.
\item Use tables and tree diagrams to compute conditional probabilities.
\item Explain the rationale behind Bayes’ theorem, and use it to compute conditional probabilities.
\item Compute probabilities in a range of settings.
\end{itemize}

\subsection{Before the session}
Students work through a directed set of materials, which may include reading, watching instructional  videos, quizzes, exercises,  among other things.

\subsection{During the session: multiple choice quiz}
Students complete a short multiple choice quiz individually, answering a range of questions on probability which might include both theoretical questions such as asking students to apply their judgement on whether two events are independent, through to computing probabilities.

Once the test is complete and submitted, students join their team and answer the same multiple choice quiz. This time each question can be discussed within the team and the group must decide on their joint final answer for each question for submission. Results are available immediately after the team multiple choice quiz, and students can argue their case with the course leader if they think they deserve a higher mark than the one they received. In statistics, this may be because questions or the choice of answers were poorly worded and thus caused confusion. 

As the results of the test are available immediately, the course leader can identify any common misconceptions or errors. A very brief lecture follows to clarify these.

\subsection{During the session: working on the problem}

An example problem in this case could be the following.

\textit{Down’s syndrome is a genetic condition resulting in some level of learning disability, with around one in every 1,000 babies born having the condition. Expectant mothers can opt to take a serum screening test to assess the risk of having a baby with Down’s syndrome. The test outcome is either `positive’ or `negative’ for Down’s syndrome.}

\textit{As with the majority of medical tests, the test isn’t 100\% accurate. From extensive research, it is known that the test is able to detect Down’s syndrome, when the baby has Down’s syndrome, in about 85\% of cases. Conversely, when the baby doesn’t have Down’s syndrome the test identifies this in about 96\% of cases.}

\textit{A pregnant woman receives a positive test for Down’s syndrome, and asks you for advice on how likely the test is to be correct. What are the chances that her baby has Down’s syndrome?}

\textbf{Possible answers:} about 85.00\%, about 2.08\%, about 0.10\%, about 0.089\% 

\textbf{Correct answer:} about 2.08\%

This problem requires students to identify the required probability from a text description, translate the given information into appropriate probabilities, and manipulate the (indirectly) given probabilities in a non-trivial manner to compute the final probability. The work involved means that it is very difficult to think of a way of splitting the work between group members: each step above depends on information from the previous step. 

The three incorrect answers are deliberately given as `common misconceptions': 85\% is the usual prosecutor's fallacy, 0.1\% is the prevalence of Down's syndrome without accounting for the additional information from a positive test, and 0.089\% represents the situation where the calculation of the probability that a test is positive is incorrect (computed without taking the complement of the specificity). It is useful that all answers have a basis in the numbers given here; if not then students won't arrive at that particular answer and so unless they are merely guessing it is of no use. 

\section{Helpful apps, websites, and technologies}\label{sec:technologies}

This section lists some resources which can be used by both lecturers and students to make sessions more interactive.

\begin{itemize}
\item General advice:
\begin{itemize}
    \item Collaboration of TBL practitioners (the Team Based Learning Collaborative) which makes example teaching material available online (\url{http://www.teambasedlearning.org/}).
    \end{itemize}
\end{itemize}
\begin{itemize}
\item Online polls and quizzes:
Students in groups could use these resources to aid each other's learning. For example, in PBL the note taker of the group could maintain the notes of the session in an interactive Padlet page instead of taking notes on a whiteboard.
\begin{itemize}
    \item Sli.do \url{https://www.sli.do/}: website to create audience polls;
    \item Kahoot \url{https://kahoot.it}: website to create and run online quizzes;
    \item Turning Point: audience response system and polling software \\ \url{https://www.turningtechnologies.com/turningpoint} ;
    \item Mentimeter \url{https://www.mentimeter.com}: App and website to create interactive presentations;
    \item Wooclap \url{https://www.wooclap.com}: create and run online quizzes, real-time discussion boards suitable for classroom use;
    \item Padlet \url{https://en-gb.padlet.com/}: Interactive web pages with a wide range of templates including note pinboards and word clouds.
\end{itemize}
\item R packages:
\begin{itemize}
    \item There is a task view on CRAN listing R packages helpful for teaching Statistics, \\ \url{https://CRAN.R-project.org/view=TeachingStatistics} .
    \item The Bayesian task view also has a section devoted explicitly to teaching Bayesian Statistics.\\
    \url{https://CRAN.R-project.org/view=Bayesian}
    \item The learnr package \url{https://rstudio.github.io/learnr/} creates R tutorials and quizzes \citep{learnr}.
    \item The exams package \url{http://www.r-exams.org/} allows a user to create quizzes from an R script. The quizzes can be exported in various formats, such as the xml format for a moodle quiz which can be embedded into a moodle page.
    \item The Shiny runtime (\url{https://shiny.rstudio.com/}) produces web applications running R code.
\end{itemize}
\item Notebook formats:\\
The notebook formats are valuable to students because they can contain a mix of writing (using either markdown or \LaTeX\ syntax), code, and the output of the code. These documents can be worked on by a group in a collaborative environment providing say RStudio server.
\begin{itemize}
  \item R Markdown Notebooks: RStudio (\url{https://rstudio.com}) provide the \verb|.nb.html| format in which the cells are active within an RStudio session. These files can also be viewed in a web browser, at which point the cells are no longer active but can still be viewed.
  \item Jupyter (formerly Ipython) notebooks, \url{https://jupyter.org/}: These notebooks allow users to distribute html documents in which the cells of the notebook execute analyses if the user has the appropriate kernel installed. If the kernel is not installed the cells cannot be executed but the documents can still be viewed in a browser.
\end{itemize}

\item Presentation and document formats:\\
Tools for creating attractive slides or documents are useful for course lecturers, but also for students if their tasks include submitting or presenting work.  
\begin{itemize}
  \item ioslides - creates html slides with interactive content, e.g. graphics. These can be produced from RMarkdown files.
  \item Prezi \url{https://prezi.com} - creates attractive  presentations which don't follow the traditional slide format.
  \item Microsoft Sway: An application to produce interactive reports and presentations. \url{https://sway.office.com/my}
  \item \LaTeX\ Beamer (the German for overhead projector). A popular modern Beamer theme is the Metropolis theme \\ \url{https://github.com/matze/mtheme}.
  \item Overleaf \url{https://www.overleaf.com}: Students can work collaboratively on \LaTeX{} documents (including reports and Beamer presentations).
\end{itemize}
\end{itemize}

\end{document}